\documentclass[preprint,showpacs,preprintnumbers,amsmath,amssymb]{revtex4}

\usepackage{graphicx}
\usepackage{amsmath, amssymb}

\begin{document}

\title{Electronic Structure of Heavy Fermion System CePt$_2$In$_7$ from Angle-Resolved Photoemission Spectroscopy}
\author{Bing Shen$^{1,2,\dag}$, Li Yu$^{1,\dag,*}$, Kai Liu$^{3,\dag}$, Shoupeng Lyu$^{1,2}$, Xiaowen Jia$^{1,4}$, E. D. Bauer$^{5}$, J. D. Thompson$^{5}$,  Yan Zhang$^{1,2}$, Chenlu Wang$^{1,2}$, Cheng Hu$^{1,2}$, Ying Ding$^{1,2}$, Xuan Sun$^{1,2}$, Yong Hu$^{1,2}$, Jing Liu$^{1,2}$, Qiang Gao$^{1,2}$,  Lin Zhao$^{1}$,  Guodong Liu$^{1}$, Zuyan Xu$^{6}$, Chuangtian Chen$^{6}$, Zhong-Yi Lu$^{3,*}$ and X. J. Zhou$^{1,2,7,*}$}
\affiliation{
\\$^{1}$National Laboratory for Superconductivity, Beijing National Laboratory for Condensed Matter Physics, Institute of Physics, Chinese Academy of Sciences, Beijing 100190, China
\\$^{2}$University of Chinese Academy of Sciences, Beijing 100049, China
\\$^{3}$Department of Physics and Beijing Key Laboratory of Opto-electronic Functional Materials $\&$ Micro-nano Devices, Renmin University of China, Beijing 100872, China
\\$^{4}$Military Transportation University, Tianjin 300161, China
\\$^{5}$Los Alamos National Laboratory, Los Alamos, New Mexico 87545, USA
\\$^{6}$Technical Institute of Physics and Chemistry, Chinese Academy of Sciences, Beijing 100080, China
\\$^{7}$Collaborative Innovation Center of Quantum Matter, Beijing 100871, China
\\$^{\dag}$These people contribute equally to the present work.
\\$^{*}$Corresponding authors: li.yu@iphy.ac.cn, zlu@ruc.edu.cn,  XJZhou@aphy.iphy.ac.cn.
}
\date{\today}

\begin{abstract}
We have carried out high-resolution angle-resolved photoemission measurements on the Ce-based heavy fermion compound CePt$_2$In$_7$ that exhibits stronger two-dimensional character than the prototypical heavy fermion system CeCoIn$_5$.  Multiple Fermi surface sheets and a complex band structure are clearly resolved.  We have also performed detailed band structure calculations on CePt$_2$In$_7$. The good agreement found between our measurements and the calculations suggests that the band renormalization effect is rather weak in CePt$_2$In$_7$.  A comparison of the common features of the electronic structure of CePt$_2$In$_7$ and CeCoIn$_5$ indicates  that CeCoIn$_5$ shows a much stronger band renormalization effect than CePt$_2$In$_7$. These results provide new information for understanding the heavy fermion behaviors and unconventional superconductivity in Ce-based heavy fermion systems.

\end{abstract}


\maketitle

\newpage

\section{Introduction}

The heavy fermion systems have been a fertile playground for investigating strong correlation, quantum criticality and unconventional superconductivity in condensed matter physics\cite{StewartReview,GegenwartReview,ThompsonReview}.  They share many characteristics of correlated d-electron high-T$_c$ cuprates and the iron-based superconductors, including the interplay between antiferromagnetism (AFM) and unconventional superconductivity\cite{Bauer2010PRB} . The origin of superconductivity in heavy fermion systems  remains an open issue.  Previous studies on the family of heavy-fermion Ce$_m$M$_n$In$_{3m+2n}$  (M=Co, Rh and Ir) compounds revealed some key ingredients in dictating their superconducting properties. One main factor is the dimensionality since  T$_c$ is found to change from 0.2 K in cubic CeIn$_3$,  to above 0.4 K in Ce$_2$MIn$_8$\cite{Kaczorowski2009, Thompson2003},  and to over 2 K in the quasi-2D CeMIn$_5$\cite{Petrovic2001, Sarrao2002, Hegger2000, Park2006, Thompson2006} when the system becomes more two-dimensional.  On the other hand, the characteristic hybridization strength between f-electrons and spd-conduction electrons also plays an important role. For example, earlier dynamical mean field theory calculations\cite{Haule2010} suggested that the f/spd hybridization can drive the CeMIn$_5$ system from a weakly hybridized antiferromagnetic limit (M=Rh) to an almost optimally hybridized superconductor with T$_c$=2.3 K (M=Co) and finally the over-hybridized paramagnetic superconductor T$_c$=0.4 K (M=Ir).  Moreover, most of the Ce$_m$M$_n$In$_{3m+2n}$ compounds exhibit a T$_c$ dome of superconductivity, which emerges around the so-called antiferromagnetic Quantum Critical Point (QCP) where the N\'{e}el temperature is suppressed to absolute zero temperature. This suggests that the fluctuation of the AFM quantum criticality may play the same important role as the phonon in the BCS superconductors, mediating the Cooper pairing in the heavy fermion systems. It is found that both the dimensionality and hybridization are strongly involved in the QCP region\cite{Bauer2010PRB, Sidorov2013}.  All these results indicate that tuning of dimensionality and hybridization  is essential in manipulating  unconventional superconductivity and revealing its microscopic mechanism  in these heavy fermion systems.


The recently discovered heavy fermion antiferromagnetic CePt$_2$In$_7$ is so far the sole member of the Ce$_m$M$_n$In$_{3m+2n}$ family with m=1 and n=2\cite{Kurenbaeva2008, Bauer2010IOP}. It exhibits some typical heavy-fermion properties like Kondo lattice coherence around 100 K, and large value of specific heat over temperature C/T$\sim$450 mJ/(mol$\cdot$K$^2$) above the T$_N$ indicating a big enhancement of the effective mass m*\cite{Altarawneh2011,YKrupko}.  The application of pressure gradually reduces the N\'{e}el temperature in CePt$_2$In$_7$,  giving rise to superconductivity and a quantum critical point at P$_c$ = 3.2--3.5 GPa\cite{Bauer2010PRB,Sidorov2013,SKurahashi}. Its crystal structure is composed of CeIn$_3$ layers separated by PtIn$_2$ layers along the {\it c} axis of the tetragonal unit cell  with lattice constants of a=4.611(1){\AA} and c=21.647(3){\AA }(Fig. 1c), and crystallizes in the body-centered I4/mmm space group\cite{Tobash2012,TKlimczuk}.  Compared with CeIn$_3$ (m=1, n=0, Fig. 1a) and CeMIn$_5$ (m=1, n=1, Fig. 1b) in the Ce$_m$M$_n$In$_{3m+2n}$ family, the distance between the two adjacent CeIn$_3$ building blocks in CePt$_2$In$_7$ (m=1, n=2, Fig. 1c) is greatly enhanced and the f/d hybridization is reduced\cite{Bauer2010PRB, Sidorov2013}, giving rise to more two-dimensional characteristics\cite{Tobash2012,Altarawneh2011,apRoberts2010}.  CePt$_2$In$_7$ therefore provides a  new platform to investigate the effect of reduced dimensionality, hybridization, and quantum critical fluctuations within the same family. There have been efforts to investigate the low-energy electronic structure of CePt$_2$In$_7$, for example, by quantum oscillation measurements\cite{Altarawneh2011}.  However, to the best of our knowledge,  direct measurement of its electronic structure from angle-resolved photoemission spectroscopy (ARPES)  is still lacking.

In this paper, we report the first ARPES measurements on the electronic structure of CePt$_2$In$_7$ combined with band structure calculations. We first present detailed Fermi surface and band structrue calculations of CePt$_2$In$_7$ by considering the spin-orbit-coupling, electron correlation, orbital characteristics of bands, and three-dimensional k$_z$ effect. Then we show our high resolution ARPES measurement on CePt$_2$In$_7$ under different polarization geometries over a wide energy range. Multiple Fermi surface sheets and complex band structure are clearly resolved. By comparing our measurements with band structure calculations, we find that most of the electronic features we observed can be accounted for by the band structure calculations although obvious discrepancies also exist. We also compare the electronic structure of CePt$_2$In$_7$ with CeCoIn$_5$\cite{Jia2011} to reveal the effect of dimensionality reduction on the electronic structure.  Our results provide new information for further investigations of heavy fermion behaviors and unconventional superconductivity in Ce-based heavy fermion systems.

\section{Experimental Methods}

Single crystals of CePt$_2$In$_7$ were grown from In-flux described in detail before\cite{Tobash2012}. We characterized the samples by carrying out x-ray diffraction (XRD) (Fig. 1d) and Laue diffraction (Fig. 1e) measurements. The results  indicate the single crystal nature of the CePt$_2$In$_7$ sample; the sharp lines in XRD pattern (Fig. 1d) and clear spots in the Laue pattern (Fig. 1e) also indicate a relatively high quality of the samples. The single crystals of CePt$_2$In$_7$ are cleavable, as evidenced  by the angle-resolved band structure shown below. However, the cleaved surface (inset photo in Fig. 1d) is a little bit rough that may limit  the data quality of our ARPES measurement on this system.

High resolution angle-resolved photoemission measurements were carried out by using our lab ARPES system equipped with a Scienta R4000 electron energy analyzer and a low temperature cryostat with 6-degree of motions\cite{GDLiu}. Here we used helium discharge lamp as the light source which can provide a photon energy of {\it h}$\upsilon$= 21.218 eV (helium I).  The spot size is $\sim$0.5 mm.  The energy resolution was set at 5$\sim$10 meV and the angular resolution was $\sim$0.3 degree. The Fermi level was referenced by measuring on a clean polycrystalline gold that is electrically connected to the sample.  The CePt$_2$In$_7$ crystal was cleaved {\it in situ} and measured in vacuum with a base pressure better than 5$\times$10$^{-11}$ Torr. The measurement temperature was set at 20 K that is well below the possible Kondo coherence onset temperature\cite{Bauer2010PRB,apRoberts2011} and above the antiferromagnetic transition temperature T$_N$$\sim$5.2K.

The density functional theory (DFT) calculations were carried out with the projector augmented wave (PAW) method\cite{Blochl1994,Kresse1999} as implemented in the VASP package\cite{Kresse1993,Kresse1996,Kresse1996PRB}. The generalized gradient approximation (GGA) of Perdew-Burke-Ernzerhof (PBE) type \cite{Perdew1996} was used for the exchange-correlation potential. The orbitals of Ce (5s$^2$5p$^6$4f$^1$5d$^1$6s$^2$), Pt (5d$^9$6s$^1$), and In (4d$^{10}$5s$^2$5p$^1$) were treated as valence electrons. The kinetic energy cutoff of the plane wave basis was set to be 350 eV. A 4$\times$4$\times$4 k-point mesh was employed for the Brillouin zone sampling. The Gaussian smearing with a width of 0.05 eV was adopted around the Fermi surface. Both cell parameters and internal atomic positions were allowed to relax until the forces on all atoms were smaller than 0.01 eV/{\AA }. The spin-orbital-coupling (SOC) effect was included in the calculations of band structure at equilibrium atomic positions. The maximally localized Wannier functions (MLWF)\cite{Marzari1997,Souza2001} was utilized to calculate the Fermi surfaces. The on-site Coulomb repulsion among the localized Ce 4f electrons was also checked by using the formalism (GGA+U) of Dudarev et al.\cite{Dudarev1998}.

\section{Band Structure Calculations of C\lowercase{e}P\lowercase{t}$_2$I\lowercase{n}$_7$}

We first present band structure calculation results on CePt$_2$In$_7$ to understand its basic electronic structure and set a stage for  a later comparison with our ARPES measurements. Fig. 2a shows the GGA-calculated  band structure of CePt$_2$In$_7$ along several high symmetry directions at the k$_z$=0 plane in the first Brillouin zone (Fig. 2f). Multiple bands are present in the energy range of 2 eV with several of them crossing the Fermi level. By including the spin-orbit-coupling in the calculations (Fig. 2b), splitting of bands occurs which is more significant for the bands near and above the Fermi level. We also considered the effect of electron correlations on the calculated electronic structure by including an on-site Coulomb repulsion (effective U=3.0 eV) which causes some slight change of band dispersions and lifting of band degeneracies (Fig. 2c).  It is found that the effect is stronger on the bands above the Fermi level than those below the Fermi level.   Overall the inclusion of spin-orbit-coupling and electron correlation does not alter the calculated band structure of CePt$_2$In$_7$ significantly, particularly below the Fermi level where the ARPES results are measured,  so we will focus on the GGA calculated results hereafter. Fig. 2d shows the GGA calculated band structure with the orbital character marked. It is clear that the low energy excitations are dominated by the Ce 4f and In 5p orbitals while the contribution from Pt 5d orbitals  is mainly away from the Fermi level. The bands that cross the Fermi level consist of either mainly Ce 4f orbitals, or mainly In 5p orbitals, or hybridized Ce 4f and In 5p orbitals. The calculated density-of-states (DOS) (Fig. 2e) is consistent with the orbital decomposition of bands (Fig. 2d) which shows that the main spectral weight above the Fermi level is dominated by Ce 4f orbitals.

In order to gain insight on the dimensionality of the electronic structure in CePt$_2$In$_7$, and to facilitate a direct comparison with our ARPES results that are measured at a particular photon energy corresponding to a momentum cut in a particular k$_z$ momentum plane, we performed detailed band structure calculations at different k$_z$ planes. Fig. 3 shows the calculated band structure of CePt$_2$In$_7$ at different k$_z$s varying from the basal k$_z$=0 plane (Fig. 3a) all the way to the k$_z$=$\pi$/c plane (Fig. 3f). Significant variation of the band structure with k$_z$ is evident.  Correspondingly, Fig. 4 shows detailed two-dimensional Fermi surface topology at different k$_z$s ranging from k$_z$=0 (Fig. 4a) to k$_z$=$\pi$/c (Fig. 4i).  The Fermi surface topology also shows an obvious change with k$_z$.  We note that  the large electron-like Fermi surface contour around M(A) point is relatively robust  against the k$_z$ variation, but for the  G(Z) point, the Fermi surface topology change is strong and complicated. These results  show a relatively two-dimensional nature of the electronic structure near M(A) point and strong  three-dimensional characteristics in other momentum space for CePt$_2$In$_7$.  The three-dimensionality of electronic structure for CePt$_2$In$_7$ is obvious even though its two-dimensionality is expected to be enhanced when compared with CeCoIn$_5$.

Figure 5 depicts the calculated three-dimensional Fermi surface of CePt$_2$In$_7$.  In order to see the complex Fermi surface clearly, we show individual Fermi surface sheets in Fig. 5(a-d),  before combining them into an overall Fermi surface (Fig. 5e). Here different color-color combination stands for the front-back facet of different Fermi surface sheet in Fig. 5(a-d). It shows a cylindrical Fermi surface sheet in cyan-red centered at Brillouin zone corner with rather strong two-dimensional character (Fig. 5d). The green-purple sheet close to the cyan-red one is also centered at corner with more 3D character and it also exhibits a small pocket around X point (Fig. 5c). Furthermore, the most three-dimensional-like Fermi surface sheets in yellow-blue and red-cyan are shown up in the zone center region and the X point region. From these results, we find that the calculated Fermi surface in the first Brillouin zone suggests two components: one part includes several clear electron pockets centered at the zone corner (M point) with strong two dimensional characteristic and the other part is the rest of the complicated features, especially at the zone center (G point), suggesting strong k$_z$-dependence or three dimensionality. We note our results are consistent with the previous band structure calculations\cite{TKlimczuk}.

\section{ARPES results of C\lowercase{e}P\lowercase{t}$_2$I\lowercase{n}$_7$ and Discussions}

Now we present our ARPES measurement results of CePt$_2$In$_7$. Since ARPES measurements involve matrix element effects where the band intensity relies on the photoemission measurement conditions, such as photon energy, photon polarization, and orbital character of a particular band\cite{ARPESReview}, we measured CePt$_2$In$_7$ using two distinct polarization geometries in order to get complete results. Fig. 6 shows our ARPES measurement results when the polarization vector of the incident light is vertical, i.e., along the $\Gamma'-X'$ direction, while Fig. 7 shows the results when the polarization vector is 45 degrees rotated with respect to the case in Fig. 6.  In both measurements, in addition to showing the measured original data, we also show second derivative images  both with respect to energy  and with respect to momentum, in order to display the measured band structure more clearly. The combination of both second-derivative images  complement each other when there are some flat bands, or steep bands, or overlapped bands in the measured data.  As seen from Figs. 6 and 7, we resolve multiple bands, more clearly displayed in the second-derivative images.

In Fig. 6,  we show measured band structure of CePt$_2$In$_7$ along two high symmetry momentum cuts, one along $M'-X'$ cut (Fig. 6(a1-a3)) and the other along $X'-\Gamma'$ cut (Fig. 6(b1-b3)).  Along $M'-X'$ cut, five bands can be clearly resolved. Four of them cross the Fermi level and their band width varies between 1.2$\sim$1.6 eV. The fifth hole-like band lies below the Fermi level with its top at a binding energy of 0.5 eV.   Along $X'-\Gamma'$ cut, one band is clearly seen that crosses the Fermi level; the band is supposed to be symmetrical with respect to $X'$ point but it is not clear on the right side of $X'$ possibly due to matrix element effect. The other bands below the Fermi level are nearly symmetrical with respect to the $X'$ point (Fig. 6b2). The other bands also form a  complicated herring-bone-like structure (Fig. 6b3).

In Fig. 7, we show band structure of CePt$_2$In$_7$ measured along three typical high symmetry momentum cuts when the polarization vector of the incident light is 45 degree rotated with respect to the case in Fig. 6.  Multiple band structure are clearly observed. Their clear symmetry with respect to high symmetry points $M'$,  $\Gamma'$ and $X'$ indicates the reliability of the measurements, and excludes the possibility that some of the observed band(s) may come from different facets of the imperfectly-cleaved surface.  Around the $M'$ point (Fig. 7(a1-a3)), at least five bands are clearly observed within the energy range measured. The bottom of the top three bands marked as M1, M2 and M3 in Fig. 7a2 lies near binding energies of 1.2 eV, 1.6 eV and 2.0eV,  respectively.  Close examination of bands in Fig. 7a2 and 7a3, and comparing with the bands near $M'$ in Fig. 6(a2),  there is possibly an extra band lying in between M1 and M2 bands, as seen more obviously in Fig. 7a2 marked by a dashed line.  Therefore, the topmost M1 crosses the Fermi level, forming an electron-like band with a Fermi momentum of $\sim$0.49$\pi$/a,  while the M2 band has a Fermi momentum  0.74$\pi$/a that is beyond our measurement momentum window, the  possible band in between M1 and M2 crosses the Fermi level at $\sim$0.65$\pi$/a (Fig. 7a2 and 7a3).   The bands deeper than 2 eV binding energy are usually related to the localized part of Ce f-electron bands\cite{Haule2010} and Pt-5d band according to earlier and present theoretical results\cite{Bauer2010PRB}.

Figure 7(b1-b3) show the measured band structure of CePt$_2$In$_7$ along $M'-\Gamma'$ momentum cut which consist of bands in three momentum regions: near $M'$, near $\Gamma'$ and somewhere in the middle between $M'$ and $\Gamma'$.  Around $M'$ point, two electron-like bands with the band bottom at binding energies of $\sim$1.2 eV and $\sim$1.6 eV are clearly observed; these two bands are consistent with the M1 and M2 bands observed in Fig. 7(a1-a3).  Near $\Gamma'$ point, five bands are clearly identified, labelled as $\Gamma1$ to $\Gamma5$ in Fig. 7b2 and 7b3. The $\Gamma$1 band is an electron-like band crossing the Fermi level with its bottom at a binding energy of $\sim$0.25 eV and with a Fermi momentum at  $\sim$0.13$\pi/a$. The $\Gamma$2 band is a flat band at a binding energy of $\sim$0.6 eV, which is consistent with that observed in Fig. 6b2.  It merges into broad band continuum in the middle momentum region between the $M'$ and $\Gamma'$ points. The $\Gamma$3 band is a hole-like band with its top at a binding energy of $\sim$0.66 eV. Right below is the $\Gamma$4 band that is also a hole-like band with its top at a binding energy of $\sim$1.0 eV. The $\Gamma$5 band lies near 1.7 eV binding energy.  In the middle between $M'$ and $\Gamma'$, there is a vertical strong spectral patch near -0.5 $\pi/a$ (Fig. 7b3). Near the Fermi level, there seems to be two bands crossing the Fermi level, marked as $\alpha$1 and $\alpha$2 in Fig. 7b3. This patch is reminiscent to that found in Fig. 6b3, also somewhere in the middle between $X'$ and $\Gamma'$ points.  Figure 7(c1-c3) show the measured band structure crossing $X'$ point. Here there is an obvious hole-like band that crosses the Fermi level with a Fermi momentum of $\sim$0.13$\pi/a$.  This is consistent with the observation of the similar hole-like band in Fig. 6a2 around $X'$ where the momentum cut is along $M'$-$X'$ direction. One can also see signatures of hole-like bands with their tops near a bind energy of 0.5 eV, 0.8 and 1.2 eV.


Figure 8 shows the Fermi surface of CePt$_2$In$_7$ measured under two different polarization geometries, and their comparison with the calculated Fermi surface.  Because of relatively weak and diffuse signal near the Fermi level possibly due to rough sample surface, the spectral intensity shown here is obtained by integrating over a 100 meV energy window near the Fermi level. Even with different polarization geometries, some features are rather robust in both measurements, including the electron-like Fermi surface sheets around $M'$ point, features surrounding the $X'$ point, and diamond-shaped pattern surrounding the $\Gamma'$ point.  The Fermi momenta and deduced Fermi surface from high resolution band structure measurements shown in Fig. 6 and Fig. 7 are also plotted onto the Fermi surface mappings.  The Fermi surface mappings shown here, combined with the band structure measurements in Fig. 6 and Fig. 7, allows us to identify clearly four Fermi surface sheets:  two electron-like Fermi surface sheets  surrounding the $M'$ points,  one hole-like Fermi surface sheet surrounding the $X'$ point, and one small electron-like Fermi surface sheet surrounding the $\Gamma'$ point.  These four Fermi surface sheets clearly identified show a good agreement with the ones in the band structure calculations (Fig. 8c).  According to GGA calculation results for CePt$_2$In$_7$ shown above, the circular Fermi surface sheets  surrounding the $M'$  zone corner have strong two-dimensional character  while the rest sheets have more  three-dimensional character.

We now have a direct comparison of the measured band structure and the calculated results of CePt$_2$In$_7$.  In the calculations, the Ce-4f electron is considered as itinerant and it certainly plays a major role on the low-energy electronic structure. Quantum oscillation measurements\cite{Altarawneh2011} suggest that the f-electrons contribute to the Fermi surface in CePt$_2$In$_7$.  The band structure measured along high-symmetry momentum cuts is summarized in Fig. 9 as the original data (Fig. 9a),  EDC-second derivative (EDC-energy distribution curve, Fig. 9b) and MDC-second derivative (MDC-momentum distribution curve, Fig. 9c) images in order to show complete information.  The calculated bands are overlaid onto Fig. 9b and 9c for a direct comparison. Because our results are measured using one particular photon energy (21.2 eV in the present paper), it corresponds to a particular k$_z$ momentum plane in a three-dimensional Brillouin zone. For CePt$_2$In$_7$ with strong three-dimensionality of its electronic structure, by comparing our measured band structure (Figs. 6, 7 and 9) and Fermi surface (Fig. 8) with those calculated results (Figs. 3 and 4) , we find that the calculated electronic structure with k$_z$$\sim$0.8$\pi/c$ show the best agreement with our experimental results. So we chose the Fermi surface (Fig. 8c) and band structure at k$_z$=0.8$\pi/c$ for comparison.

Overall the measured band structure of CePt$_2$In$_7$ shows surprisingly good agreement with the band structure calculations, in terms of the number of bands observed, the bandwidth, and even the quantitative band dispersions.  Along $X'$-$M'$ momentum cut, the four bands crossing the Fermi level show rather good agreement between measurements and calculations. Two other calculated bands can also find their signatures in the measured data. At $\Gamma'$ point, five sets of bands are observed, in good agreement with the calculated results, including the relatively flat band near the 0.6 eV binding energy (Fig. 9b). The three bands clearly shown up in the $\Gamma'$-$X'$ momentum cut can also find their counterparts in the calculations. Along $M'$-$\Gamma'$ cut, a good agreement exists as for the four bands near $M'$ although the number of the observed bands seems less than the calculated ones.  The most obvious difference between the measurements and calculations lies in the two nearly vertical spectral patched on either side of the $\Gamma'$ point (Fig. 9c).  It  remains to be investigated whether fine tuning of k$_z$ or inclusion of correlation effect in the calculations reproduce these peculiar features.  The good match between the measured and calculated band structure indicates that the band renormalization effect in CePt$_2$In$_7$ is unexpectedly weak.

Finally we directly compare the electronic structure of CePt$_2$In$_7$ with that of CeCoIn$_5$\cite{Jia2011,Koitzsch2009,Koitzsch2013}, both compounds belonging to the same Ce$_m$M$_n$In$_{3m+2n}$ family and containing the same CeIn$_3$ building blocks that dictates the major low energy electronic structure.  Fig. 10 compares both the band structure and Fermi surface between these two heavy Fermion systems. Apparently CePt$_2$In$_7$ shows more bands than CeCoIn$_5$ in the energy and momentum space covered. One common feature is the existence of quasi-two-dimensional electron-like Fermi surface sheet around the $M'$ point. According to the band structure calculations (Fig. 2d), the band is mainly dominated by In 5p orbitals hybridized with slight Ce 4f orbitals. Another common feature is the observation of a section of nearly flat band near $\Gamma'$ point that lies at 0.6 eV binding energy in CePt$_2$In$_7$ (Fig. 10a) and $\sim$0.3 eV in CeCoIn$_5$ (Fig. 10b).   The most pronounced difference between CePt$_2$In$_7$ and CeCoIn$_5$ is the band width. For the electron-like band around $M'$ point, the bottom of the band lies at 1.2 eV binding energy in CePt$_2$In$_7$, but only at 0.5 eV binding energy in CeCoIn$_5$. For the band with similar orbital character, the band renormalization in CeCoIn$_5$ is 2.4 times that of CePt$_2$In$_7$. The same is true for the flat band near $\Gamma$ where it is twice deep in CePt$_2$In$_7$ that in CeCoIn$_5$. These results indicate that, with the insertion of PtIn$_2$ layers in between CeIn$_3$ building blocks to increase the distance, the electron correlation in CePt$_2$In$_7$ seems to be obviously suppressed.

\section{Summary}

In  summary, we have presented the first high resolution ARPES study on the CePt$_2$In$_7$ heavy fermion system. Clear Fermi surface and band structure are resolved. The measured results show a good agreement with the band structure calculations. It is found that the band renormalization effect in CePt$_2$In$_7$ is rather weak. It is more than two times weaker than that in CeCoIn$_5$. These results will provide information for further investigations of the heavy fermion behaviors in the Ce-based systems. It remains to be investigated why the renormalization is obviously different between CeCoIn$_5$ and CePt$_2$In$_7$ although they share similar CeIn$_3$ building blocks. Further work is needed to uncover the origin of heavy Fermion behaviors in CePt$_2$In$_7$, i.e., which band(s) are responsible for giving rise to the heavy Fermion properties.  The revelation of bands in CePt$_2$In$_7$  also paves a way for investigating the many-body effects to sort out the origin of its physical properties and the pressure-induced superconductivity.

\vspace{3mm}

\noindent {\bf Acknowledgement}\\
The ARPES experimental work is supported by the National Natural Science Foundation of China (Grant No. 11574360), the National Basic Research Program of China (Grant Nos. 2015CB921300, 2013CB921700, and 2013CB921904), and the Strategic Priority Research Program (B) of the Chinese Academy of Sciences (Grant No. XDB07020300). The calculation work is supported by the National Natural Science Foundation of China (Grant No. 91421304), the Fundamental Research Funds for the Central Universities of China, and the Research Funds of Renmin University of China (Grant Nos. 14XNLQ03 and 16XNLQ01). Computational resources are provided by the Physical Laboratory of High Performance Computing at Renmin University. The calculated Fermi surfaces were prepared with the XCRYSDEN program\cite{Kokalj2003}. Single crystal growth at Los Alamos National Laboratory was supported by the U.S. Department of Energy, Office of Basic Energy Sciences, Division of Materials Sciences and Engineering.





\renewcommand\figurename{Fig.}

\newpage

\begin{figure*}[tbp]
\begin{center}
\includegraphics[width=1.0\columnwidth,angle=0]{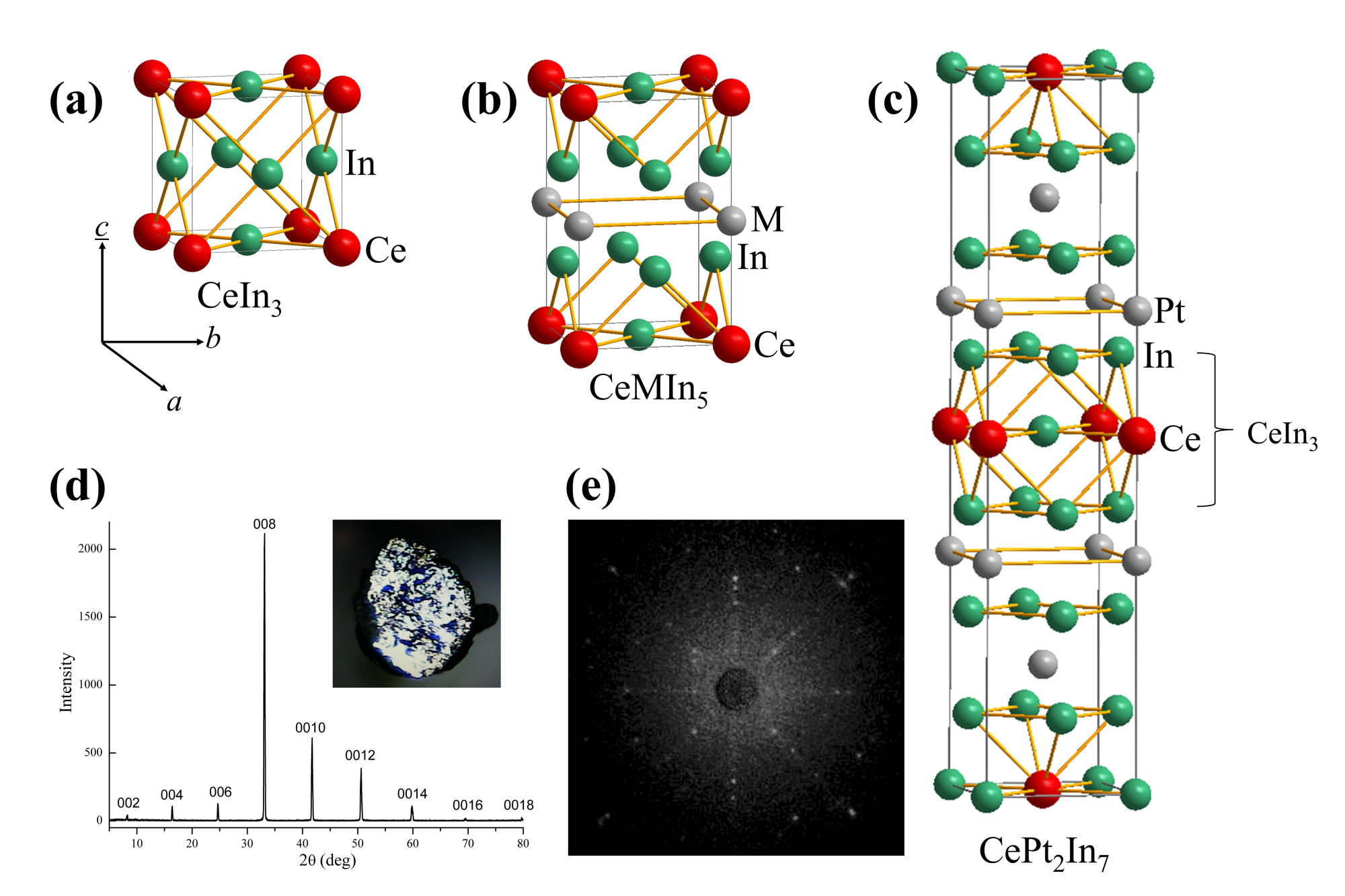}
\end{center}
\caption{\textbf{Crystal structure and characterization of CePt$_2$In$_7$.} (a) Crystal structure of CeIn$_3$. (b) Crystal structure of CeMIn$_5$ (M=Co, Rh and Ir) with a space group P4/mmm. (c) Crystal structure of CePt$_2$In$_7$ with a space group I4/mmm, formed with CeIn$_3$ layers separated by PtIn$_2$ layers. (d) X-Ray diffraction patten of our CePt$_2$In$_7$ single crystal. The inset shows the image of the  surface of a cleaved CePt$_2$In$_7$ sample with a size of 1mm$\times$0.8mm. (e) Laue diffraction image of our CePt$_2$In$_7$ sample.
}

\end{figure*}

\begin{figure*}[tbp]
\begin{center}
\includegraphics[width=1.0\columnwidth,angle=0]{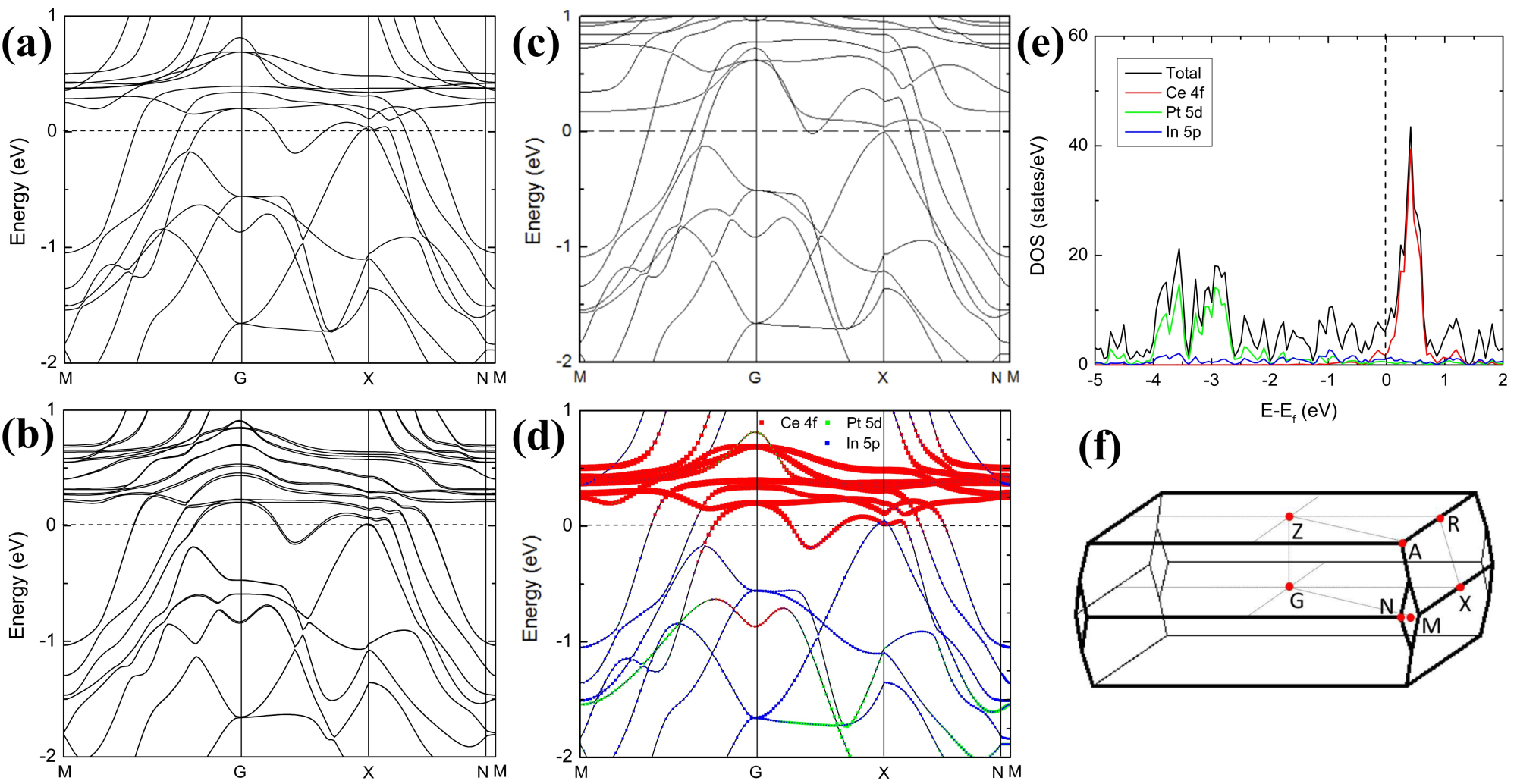}
\end{center}
\caption{Calculated band structure of CePt$_2$In$_7$ along high-symmetry directions of the Brillouin zone (BZ).  (a) Calculated results without including spin-orbital-coupling (SOC).  (b) Calculated band structure including the SOC. (c) Calculated band structure with the on-site Coulomb repulsion (effective U = 3.0 eV).  (d) Calculated band structure with the orbital characteristics marked. The red dots represent the Ce-4f orbitals, the blue dots represent the In-5p orbitals, and the green dots represent the Pt-5d orbitals. The size of the dots represents the related orbital weight.  (e) The total and partial density of states corresponding to calculations (a). (f) Bulk Brillouin zone of CePt$_2$In$_7$ with high symmetry momentum points marked. The locations of G, X, N and M points are at k$_z$=0 plane while the corresponding Z, R and A points are at k$_z$=$\pi$/\emph{c} plane.
}

\end{figure*}

\begin{figure*}[tbp]
\begin{center}
\includegraphics[width=1.0\columnwidth,angle=0]{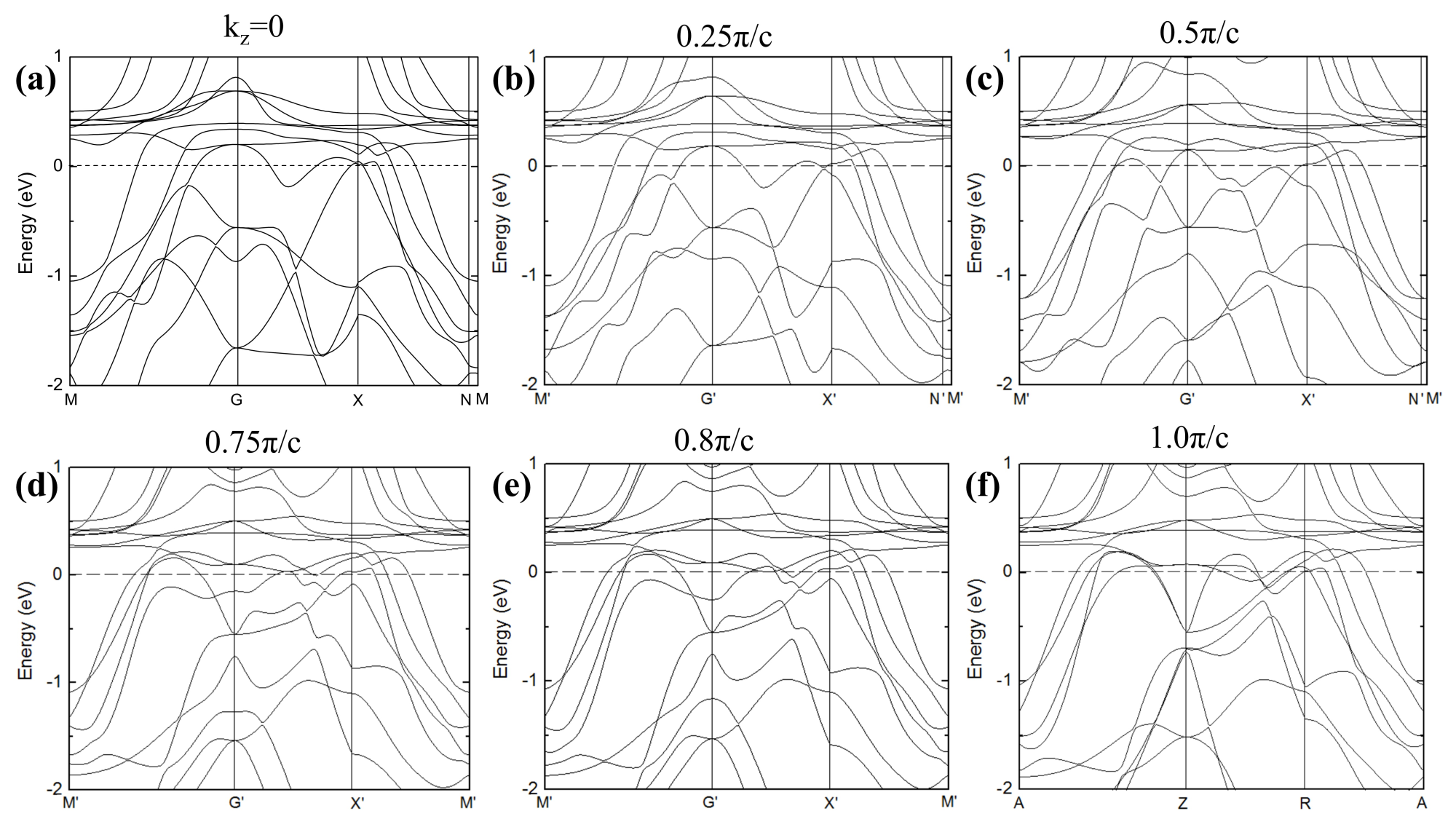}
\end{center}
\caption{Band structure of CePt$_2$In$_7$ along high-symmetry directions of the Brillouin zone at different k$_z$s of (a) 0, (b) 0.25 $\pi$/\emph{c}, (c) 0.5 $\pi$/\emph{c}, (d) 0.75 $\pi$/\emph{c}, (e) 0.8 $\pi$/\emph{c}, and (f) $\pi$/\emph{c}. \emph{c} is the lattice constant along the z direction.
}

\end{figure*}

\begin{figure*}[tbp]
\begin{center}
\includegraphics[width=1.0\columnwidth,angle=0]{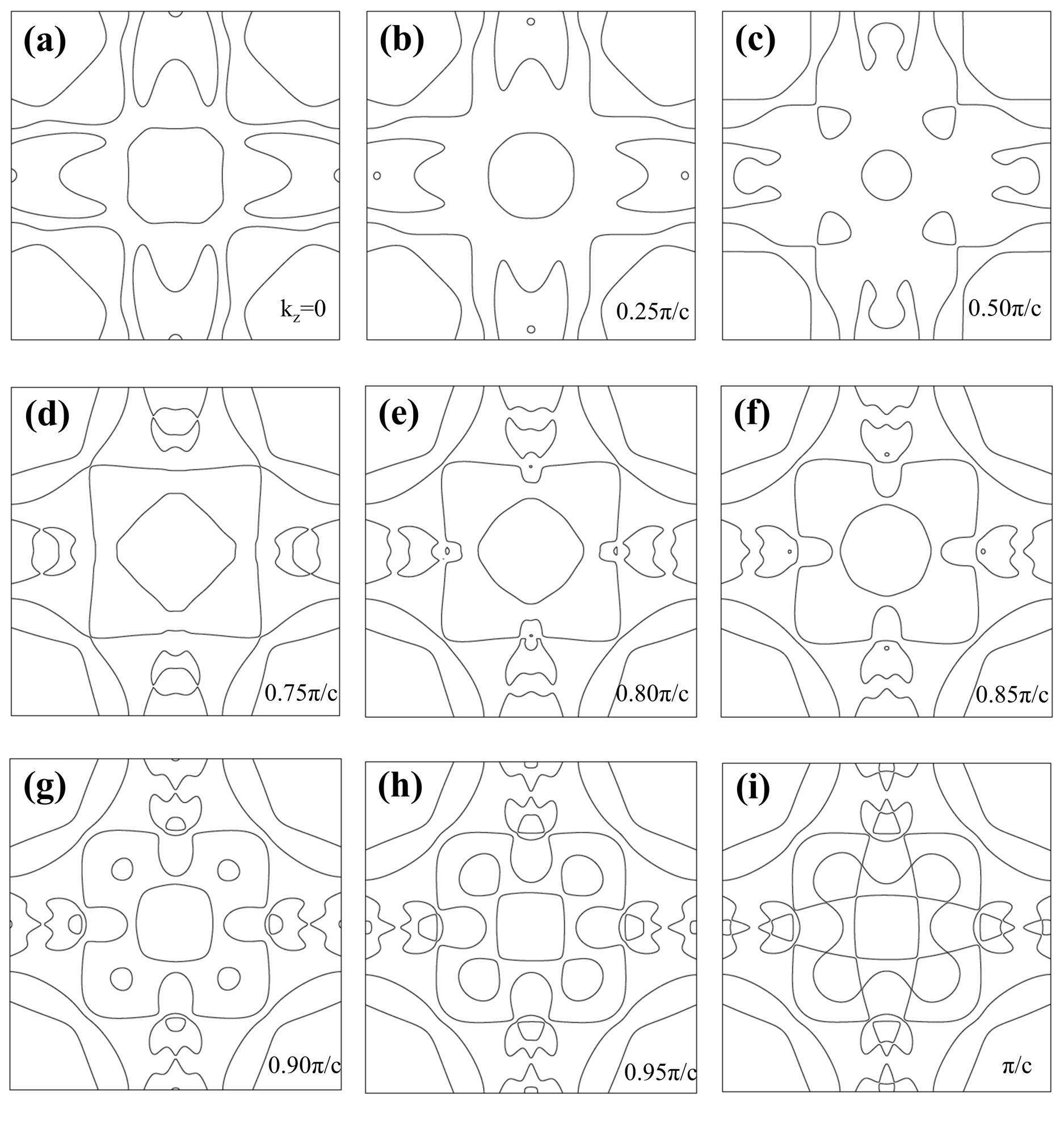}
\end{center}
\caption{Two-dimensional Fermi surface contours of CePt$_2$In$_7$ at different k$_z$s of (a) 0, (b) 0.25 $\pi$/\emph{c}, (c) 0.5 $\pi$/\emph{c}, (d) 0.75 $\pi$/\emph{c}, (e) 0.8 $\pi$/\emph{c}, (f) 0.85 $\pi$/\emph{c}, (g) 0.9 $\pi$/\emph{c}, (h) 0.95 $\pi$/\emph{c}, and (i) $\pi$/\emph{c}. \emph{c} is the lattice constant along the z direction.
}

\end{figure*}

\begin{figure*}[tbp]
\begin{center}
\includegraphics[width=1.0\columnwidth,angle=0]{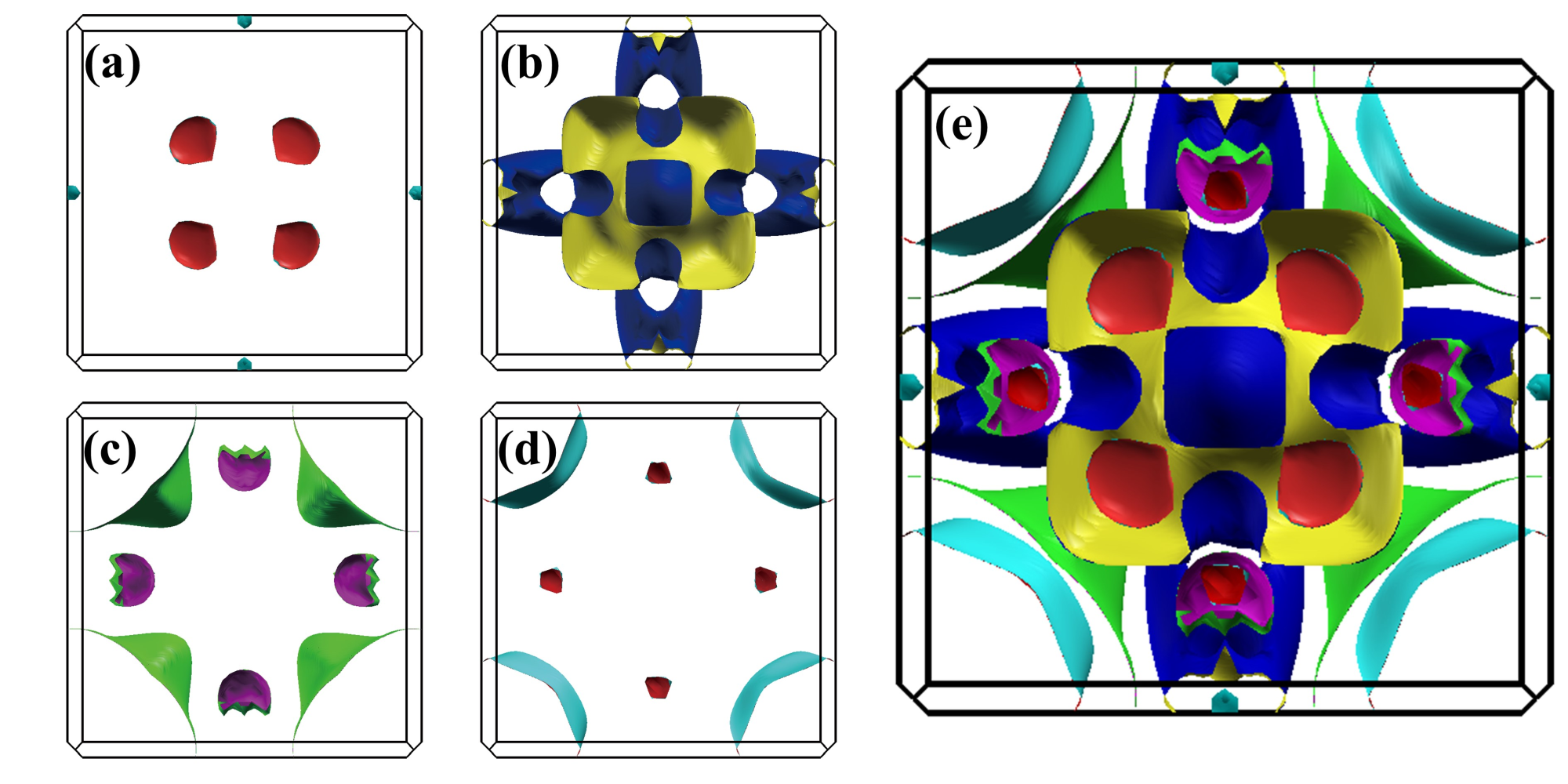}
\end{center}
\caption{Calculated Fermi surface of CePt$_2$In$_7$.  (a-d) Top views of several isolated Fermi surface sheets. (e) Merged three-dimensional Fermi surface  of CePt$_2$In$_7$.  Here different color-color combination stands for different Fermi surface sheet.
}

\end{figure*}

\begin{figure*}[tbp]
\begin{center}
\includegraphics[width=1.0\columnwidth,angle=0]{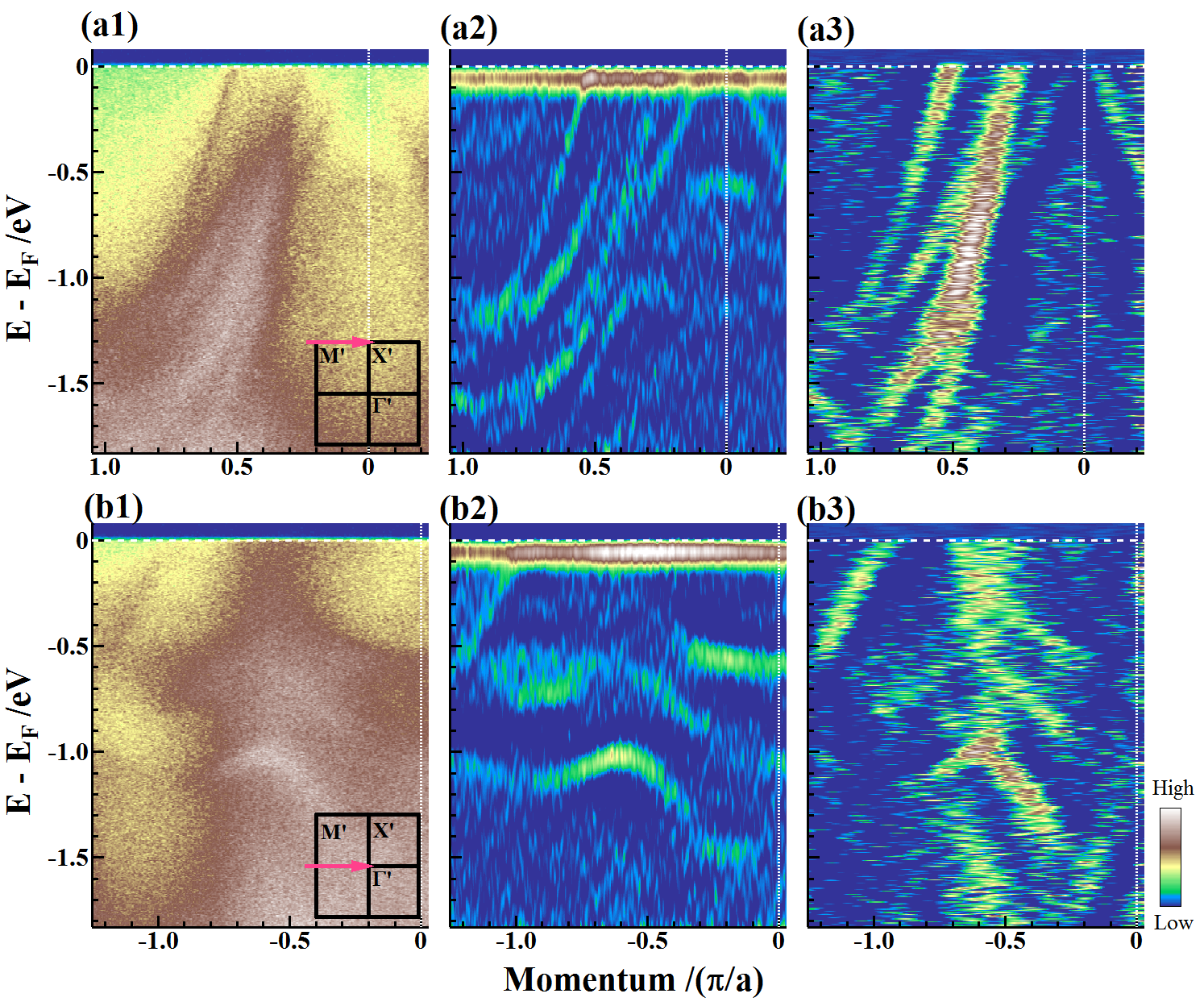}
\end{center}
\caption{\textbf{Band structure of CePt$_2$In$_7$ measured along two high-symmetry momentum cuts. } The electric field direction of the incident light is vertical, i.e, along $\Gamma'-X'$ direction. (a1-a3) Measured band structure along the $M'-X'$ direction. The location of the momentum cut  is shown by the pink arrowed line in the inset of (a1).   (b1-b3) Measured band structure along $X'-\Gamma'$ direction. The location of the momentum cut  is shown by the pink arrowed line in the inset of (b1).   (a1) and (b1) are original data.  (a2) and (b2) are  second-derivative images of the original data with respect to energy.  (a3) and (b3) are second derivative images of the original data with respect to momentum.
}

\end{figure*}

\begin{figure*}[tbp]
\begin{center}
\includegraphics[width=1.0\columnwidth,angle=0]{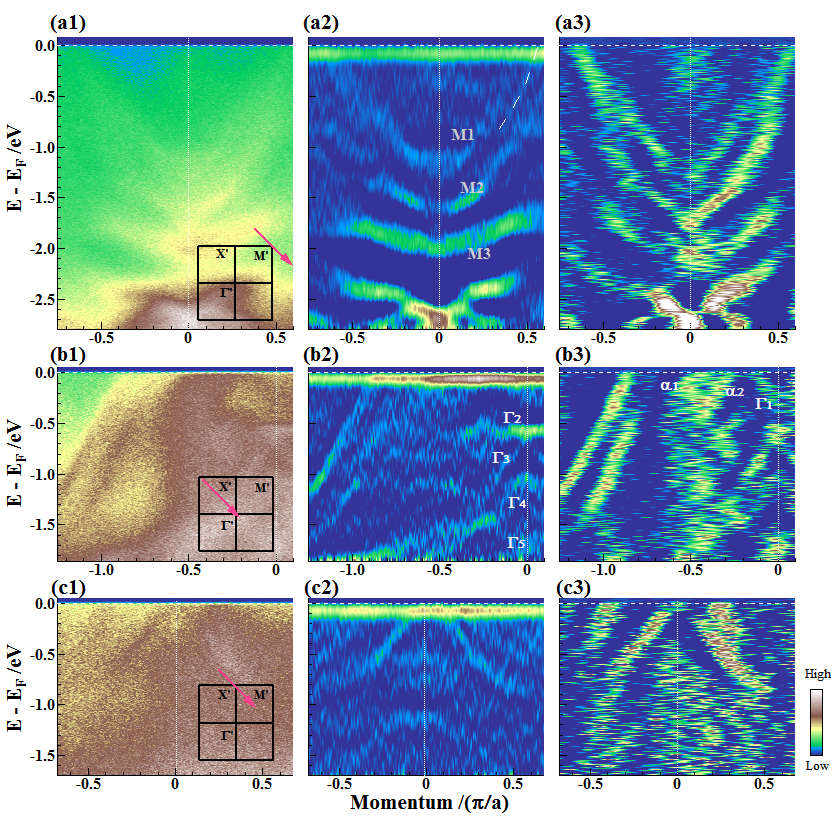}
\end{center}
\caption{\textbf{Band structure of CePt$_2$In$_7$ measured along three typical momentum cuts. }  The photon electric field direction is along $\Gamma'-M'$ diagonal direction, i.e., 45 degree rotated with respect to the case in Fig. 6. (a1-a3) Band structure measured along a momentum cut around $M'$ point. The location of the momentum cut  is shown by the pink arrowed line in the inset of (a1). The pink dashed line in Fig. 7a2 marks a possible extra band between the M1 and M2 bands.  (b1-b3) Band structure measured along $M'-\Gamma'$ direction. The location of the momentum cut  is shown by the pink arrowed line in the inset of (b1).  (c1-c3) Band structure measured along a cut crossing $\Gamma'$ point. The location of the momentum cut  is shown by the pink arrowed line in the inset of (c1).  (a1),(b1) and (c1) are original data. (a2),(b2) and (c2) are second derivative images of the original data with respect to energy. (a3),(b3) and (c3) are second derivative images of the original data with respect to momentum.
}

\end{figure*}

\begin{figure*}[tbp]
\begin{center}
\includegraphics[width=1.0\columnwidth,angle=0]{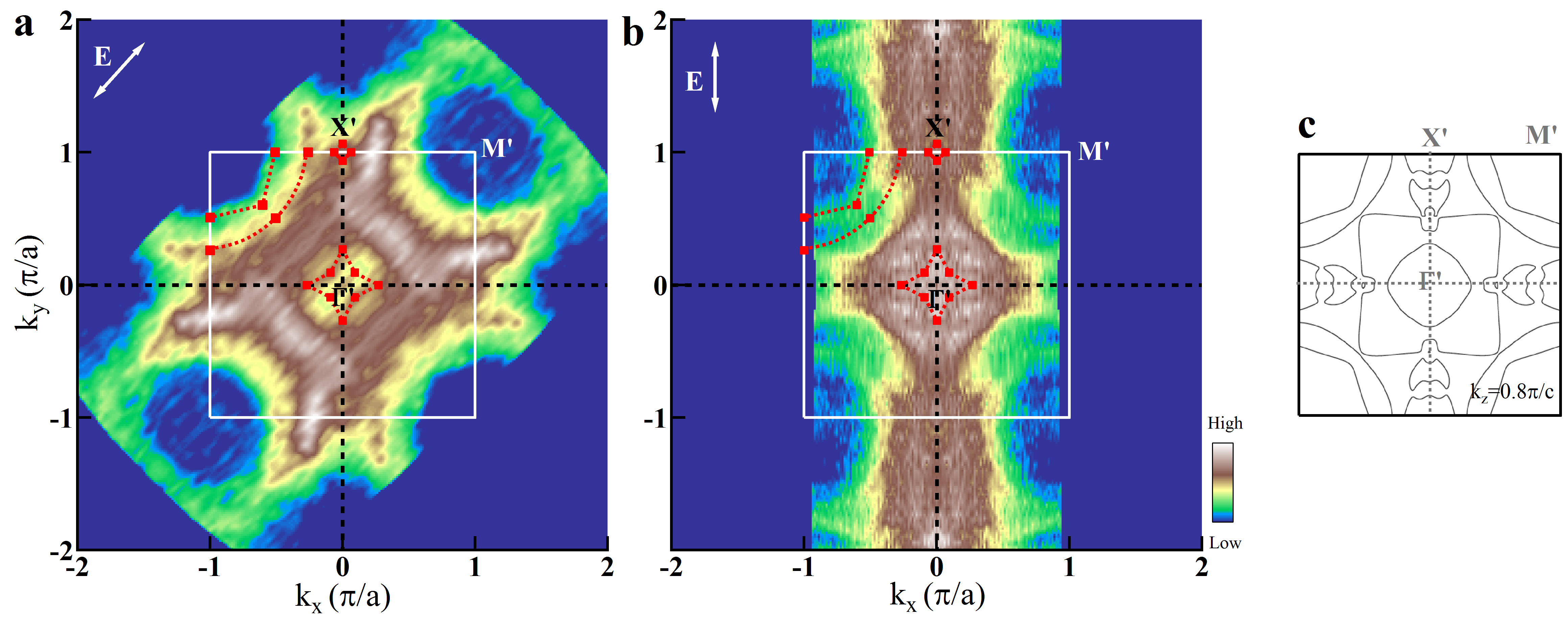}
\end{center}
\caption{\textbf{Measured Fermi surface of CePt$_2$In$_7$ and their comparison with calculated one.} (a) Fermi surface measured with the electric field direction of the incident light along the diagonal direction, as indicated by the while line in top-left corner.  Here the original data were taken in the area covered by (-2,2)-(0,0)-(2,2) region, and the full Fermi surface is obtained by symmetrizing the original data using four-fold symmetry. (b) Fermi surface measured with the electric field direction of the incident light vertical. The spectral weight is obtained by integrating over [0, 100meV] energy window near the Fermi level.  Here the original data were taken in the area covered by (-2,0)-(0,0)-(0,2) region, and the full Fermi surface is obtained by symmetrizing the original data using four-fold symmetry.  The thick white square marks the first Brillouin zone boundary. The symbols in (a) and (b) represent Fermi momenta obtained from detailed band structure measurements as shown in Figs. 6 and 7. The red lines represent Fermi surface sheets deduced from the band structure measurements in Figs. 6 and 7.  (c) Calculated Fermi surface at k$_z$=0.8 $\pi$/\emph{c} where only the first Brillouin zone is depicted.
}
\end{figure*}

\begin{figure*}[tbp]
\begin{center}
\includegraphics[width=1.0\columnwidth,angle=0]{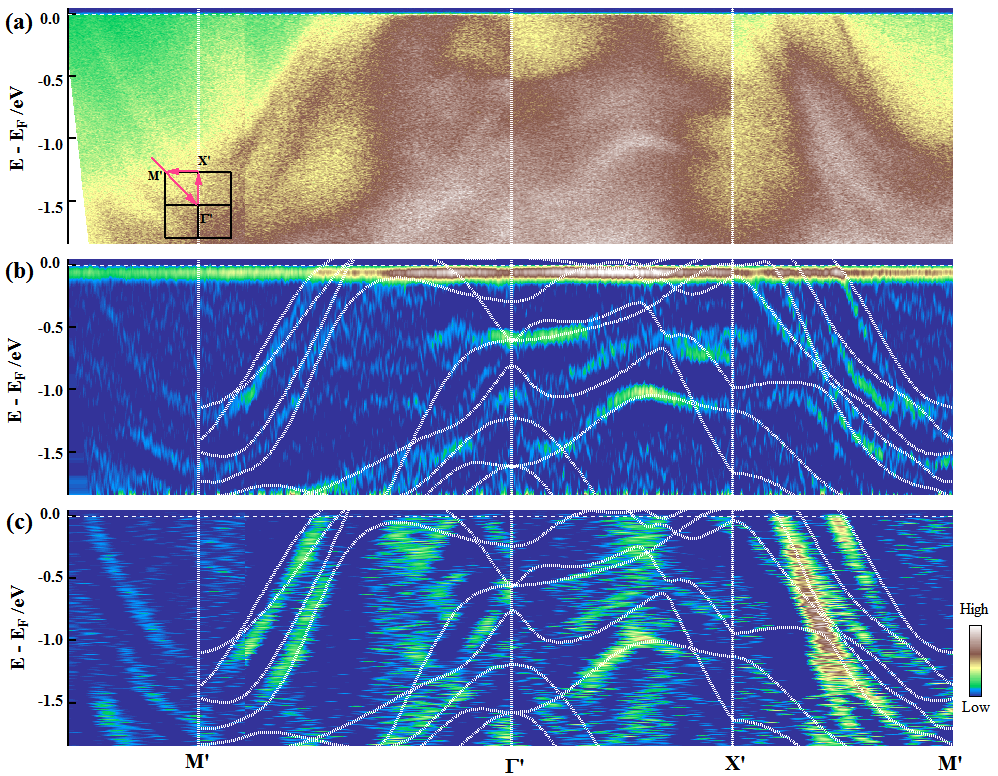}
\end{center}
\caption{\textbf{Measured band structure of CePt$_2$In$_7$ along high-symmetry directions M'-$\Gamma$'-X'-M' of the BZ and the comparison with calculated  results.} (a) Original data consisting measurements along three momentum cuts. The location of the momentum cuts are indicated by pink lines in the inset.   (b) Second derivative image of  (a) with respect to energy. (c) Second derivative image of (a)  with respect to momentum. The dashed white lines in (b) and (c) represent calculated energy bands at k$_z$=0.8 $\pi$/\emph{c}.
}
\end{figure*}

\begin{figure*}[tbp]
\begin{center}
\includegraphics[width=1.0\columnwidth,angle=0]{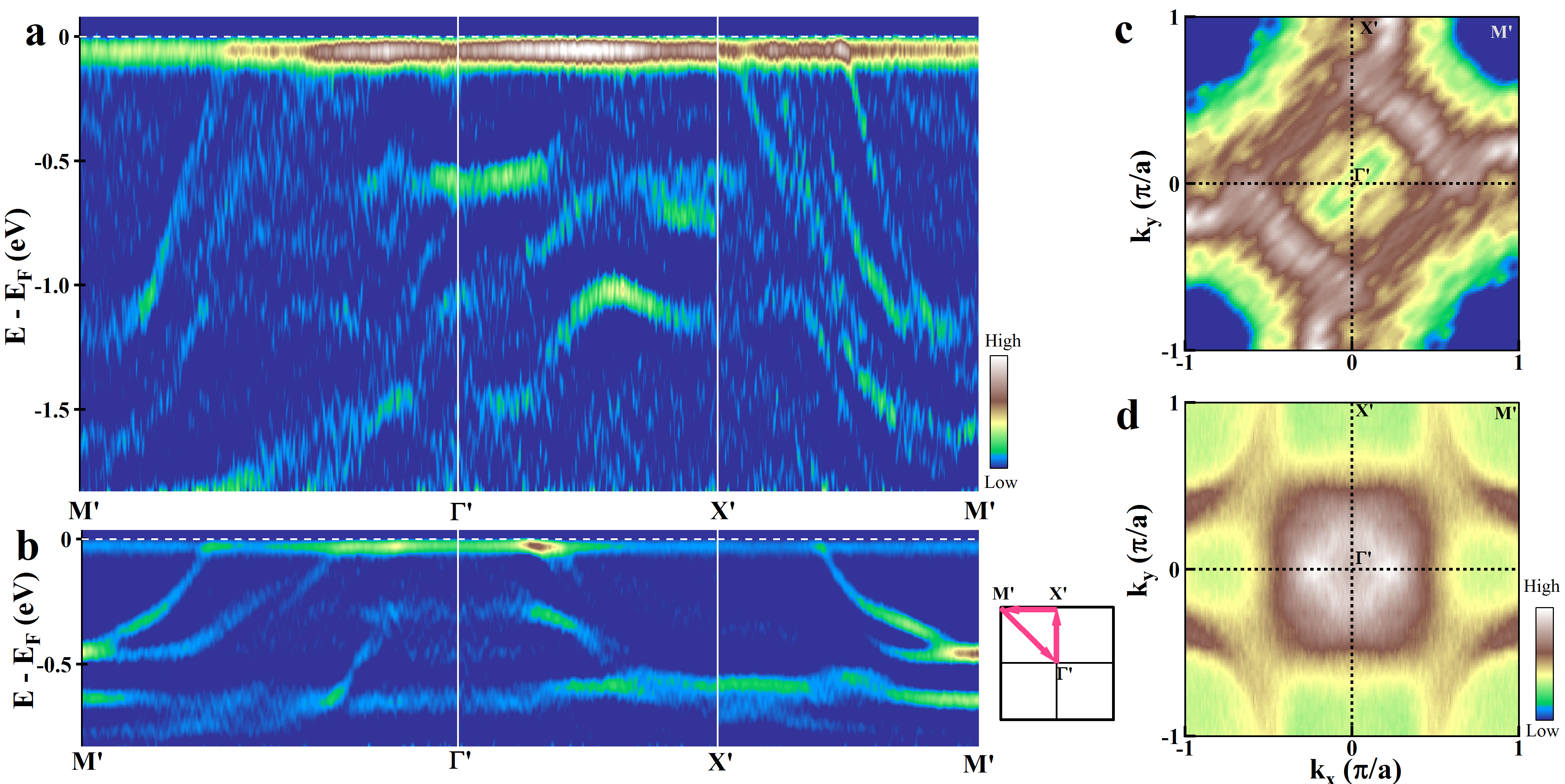}
\end{center}
\caption{\textbf{Comparison of Fermi surface  and band structure between CePt$_2$In$_7$ and CeCoIn$_5$.} (a) Measured band structure of CePt$_2$In$_7$ which is a second derivative image of the original data with respect to energy.  (b) Measured band structure of CeCoIn$_5$ which is a second derivative image of the original data with respect to energy. The location of the momentum cuts for (a) and (b) is shown by the pink lines in the inset on the right of (b).  (c) Measured Fermi surface of CePt$_2$In$_7$. (d) Measured Fermi surface of CeCoIn$_5$\cite{Jia2011}.
}
\end{figure*}


\vspace{3mm}



\begin{thebibliography}{99}
\bibitem{StewartReview} G. R. Stewart, Rev. Mod. Phys. 56, 755 (1984).
\bibitem{GegenwartReview} P. Gegenwart, Q. M. Si and F. Steglich, Nature Phys.  4, 186 (2008).
\bibitem{ThompsonReview} J. D. Thompson and Z. Fisk,  J. Phys. Soc. Jpn. 81, 011002 (2012).
\bibitem{Bauer2010PRB}E. D. Bauer, H. O. Lee, V. A. Sidorov, N. Kurita, K. Gofryk, J. X. Zhu, F. Ronning, R. Movshovich, J. D. Thompson, and Tuson Park, Phys. Rev. B 81, 180507 (2010).
\bibitem{Kaczorowski2009}D. Kaczorowski, A. P. Pikul, D. Gnida, and V. H. Tran, Phys.Rev. Lett. 103, 027003 (2009).
\bibitem{Thompson2003}J.D. Thompson, M. Nicklas, A. Bianchi, R. Movshovich, A. Llobet, W. Bao, A. Malinowski, M.F. Hundley, N.O. Moreno, P.G. Pagliuso, J.L. Sarrao, S. Nakatsuji, Z. Fisk, R. Borth, E. Lengyel, N. Oeschler, G. Sparn, and F. Steglich, Physica B 329-333, 446 (2003).
\bibitem{Petrovic2001}C. Petrovic, P. G. Pagliuso, M. F. Hundley, R. Movshovich, J. L. Sarrao, J. D. Thompson, Z. Fisk, and P. Monthoux, J. Phys.: Condens. Matter 13, L337 (2001).
\bibitem{Sarrao2002}J. L. Sarrao, L. A. Morales, J. D. Thompson, B. L. Scott, G. R. Stewart, F. Wastin, J. Rebizant, P. Boulet, E. Colineau, and G. H. Lander, Nature 420, 297 (2002).
\bibitem{Hegger2000}H. Hegger, C. Petrovic, E. G. Moshopoulou, M. F. Hundley, J. L.Sarrao, Z. Fisk, and J. D. Thompson, Phys. Rev. Lett. 84, 4986 (2000).
\bibitem{Park2006}T. Park, F. Ronning, H. Q. Yuan, M. B. Salamon, R. Movshovich, J. L. Sarrao, and J. D. Thompson, Nature 440, 65 (2006).
\bibitem{Thompson2006}J. D. Thompson, M. Nicklas, V. A. Sidorov, E. D. Bauer, R. Movshovich, N. J. Curro, and J. L. Sarrao, J. Alloys Compd. 408-412, 16 (2006).
\bibitem{Haule2010}K. Haule, C.-H. Yee, and K. Kim, Phys. Rev. B 81, 195107 (2010).
\bibitem{Sidorov2013}V. A. Sidorov, Xin Lu, T. Park, Hanoh Lee, P. H. Tobash, R. E. Baumbach, F. Ronning, E. D. Bauer, and J. D. Thompson, Phys. Rev. B 88, 020503 (2013).

\bibitem{Kurenbaeva2008}Zh M. Kurenbaeva, E. V. Murashova, Y. D. Seropegin, H. Noel, and A. Tursina, Intermetallics 16, 979 (2008).
\bibitem{Bauer2010IOP}E. D. Bauer, V. A. Sidorov, H. Lee, N. Kurita, F. Ronning, R. Movshovich, and J. D. Thompson, J. Phys.:Conf. Ser. 200, 012011 (2010).

\bibitem{Altarawneh2011}M. M. Altarawneh, N. Harrison, R. D. McDonald, F. F. Balakirev,C. H. Mielke, P. H. Tobash, J.X. Zhu, J. D. Thompson, F. Ronning, and E. D. Bauer, Phys. Rev. B 83, 081103 (2011).
\bibitem{YKrupko} Y. Krupko, A. Demuer,  S. Ota,  Y. Hirose,  R. Settai and I. Sheikin,  Phys.  Rev. B 93,  085121 (2016).  

\bibitem{SKurahashi} S. Kurahashi, S. Ota, S. Tomaru, Y. Hirose, and R. Settai, J. Phys.: Conf. Ser. 592, 012006 (2015).


\bibitem{Tobash2012}P. H. Tobash, F. Ronning, J. D. Thompson, B. L. Scott, P. J. W. Moll, B. Batlogg, and E. D. Bauer, J. Phys.: Condens. Matter 24, 015601 (2012).

\bibitem{TKlimczuk}  T.  Klimczuk, O. Walter, L.  Muechler, J. W. Krizan, F. Kinnart and R. J. Cava,  J. Phys.: Condens. Matter 26,  402201 (2014).

\bibitem{apRoberts2010}N. apRoberts-Warren, A. P. Dioguardi, A. C. Shockley, C. H. Lin, J. Crocker, P. Klavins, and N. J. Curro, Phys. Rev. B 81, 180403 (2010).


\bibitem{Jia2011}X. W. Jia, Y. Liu, L. Yu, J. F. He, L. Zhao, W. T. Zhang, H. Y. Liu, G. D. Liu, S. L. He, J. Zhang, W. Lu, Y. Wu, X. L. Dong, L. L. Sun, G. L. Wang, Y. Zhu, X. Y. Wang, Q. J. Peng, Z. M. Wang, S. J. Zhang, F. Yang, Z. Y. Xu, C. T. Chen, and X. J. Zhou, Chin. Phys. Lett. 28, 057401 (2011).

\bibitem{GDLiu}G. D. Liu, G. L. Wang, Y.  Zhu, H. B. Zhang, G. C. Zhang, X. Y. Wang, Y.  Zhou, W. T.  Zhang, H. Y.  Liu, L. Zhao, J. Q. Meng, X. L. Dong, C. T. Chen, Z. Y. XU and X. J. Zhou,  Rev. Sci. Instrum.  79, 023105 (2008).

\bibitem{apRoberts2011}N. apRoberts-Warren, A. P. Dioguardi, A. C. Shockley, C. H. Lin, J. Crocker, P. Klavins, D. Pines, Y.-F. Yang, and N. J. Curro, Phys. Rev. B 83, 060408 (2011).

\bibitem{Blochl1994}P. E. Bl\"{o}chl, Phys. Rev. B 50, 17953 (1994).
\bibitem{Kresse1999}G. Kresse and D. Joubert, Phys. Rev. B 59, 1758 (1999).
\bibitem{Kresse1993}G. Kresse and J. Hafner, Phys. Rev. B 47, 558(R) (1993).
\bibitem{Kresse1996}G. Kresse and J. Furthm\"{u}ller, Comput. Mater. Sci. 6, 15 (1996).
\bibitem{Kresse1996PRB}G. Kresse and J. Furthm\"{u}ller, Phys. Rev. B 54, 11169 (1996).
\bibitem{Perdew1996} J. P. Perdew, K. Burke and M. Ernzerhof, Phys. Rev. Lett. 77, 3865 (1996).
\bibitem{Marzari1997}N. Marzari and D. Vanderbilt, Phys. Rev. B 56, 12847 (1997).
\bibitem{Souza2001}I. Souza, N. Marzari and D. Vanderbilt, Phys. Rev. B 65, 035109 (2001).
\bibitem{Dudarev1998}S. L. Dudarev, G. A. Botton, S. Y. Savrasov, C. J. Humphreys and A. P. Sutton, Phys. Rev. B 57, 1505 (1998).

\bibitem{ARPESReview}A. Damascelli, Z. Hussain and Z. X. Shen, Rev. Mod. Phys. 75, 473 (2003).

\bibitem{Koitzsch2009}A. Koitzsch, I. Opahle, S. Elgazzar, S. V. Borisenko, J. Geck, V. B. Zabolotnyy, D. Inosov, H. Shiozawa, M. Richter, M. Knupfer, J. Fink, B. B\"{u}chner, E. D. Bauer, J. L. Sarrao, and R. Follath, Phys. Rev. B 79, 075104 (2009).
\bibitem{Koitzsch2013}A. Koitzsch, T. K. Kim, U. Treske, M. Knupfer, B. B\"{u}chner, M. Richter, I. Opahle, R. Follath, E. D. Bauer, and J. L. Sarrao, Phys. Rev. B 88,  035124 (2013).

\bibitem{Kokalj2003}A.  Kokalj, Comput. Mater. Sci. 28, 155 (2003).

\end{thebibliography}
\end{document}